\documentclass[utf8]{frontiersFPHY} 
\setcitestyle{square} 
\usepackage{url,hyperref,lineno,microtype,subcaption}
\usepackage[onehalfspacing]{setspace}

\usepackage[abs]{overpic}

\def\keyFont{\fontsize{8}{11}\helveticabold }
\def\firstAuthorLast{Kovalev and Sandhoefner} 
\def\Authors{Alexey A. Kovalev\,$^{1,*}$ and Shane Sandhoefner\,$^{1}$}

\def\Address{$^{1}$Department of Physics and Astronomy and Nebraska Center for Materials and Nanoscience, University of Nebraska, Lincoln, Nebraska 68588, USA }
\def\corrAuthor{Alexey A. Kovalev}

\def\corrEmail{alexey.kovalev@unl.edu}

\begin{document}
\onecolumn
\firstpage{1}

\title[Skyrmions and antiskyrmions in quasi-two-dimensional magnets]{Skyrmions and antiskyrmions in quasi-two-dimensional magnets} 

\author[\firstAuthorLast ]{\Authors} 
\address{} 
\correspondance{} 

\extraAuth{}

\maketitle

\begin{abstract}

\section{}
A stable skyrmion, representing the smallest realizable magnetic texture, could be an ideal element for ultra-dense magnetic memories. Here, we review recent progress in the field of skyrmionics, which is concerned with studies of tiny whirls of magnetic configurations for novel memory and logic applications, with a particular emphasis on antiskyrmions. Magnetic antiskyrmions represent analogs of skyrmions with opposite topological charge. Just like skyrmions, antiskyrmions can be stabilized by the Dzyaloshinskii-Moriya interaction, as has been demonstrated in a recent experiment. Here, we emphasize differences between skyrmions and antiskyrmions, e.g., in the context of the topological Hall effect, skyrmion Hall effect, as well as nucleation and stability. Recent progress suggests that anitskyrmions can be potentially useful for many device applications. Antiskyrmions offer advantages over skyrmions as they can be driven without the Hall-like motion, offer increased stability due to dipolar interactions, and can be realized above room temperature.

\tiny
 \keyFont{ \section{Keywords:} Dzyaloshinskii–Moriya interaction, chiral spin textures, race-track memory, topological transport, magnetic skyrmion, antiskyrmion, skyrmion Hall effect, skyrmion stability} 
\end{abstract}

\section{Introduction}
\begin{figure}[h!]
\begin{center}
\includegraphics[width=0.8\columnwidth]{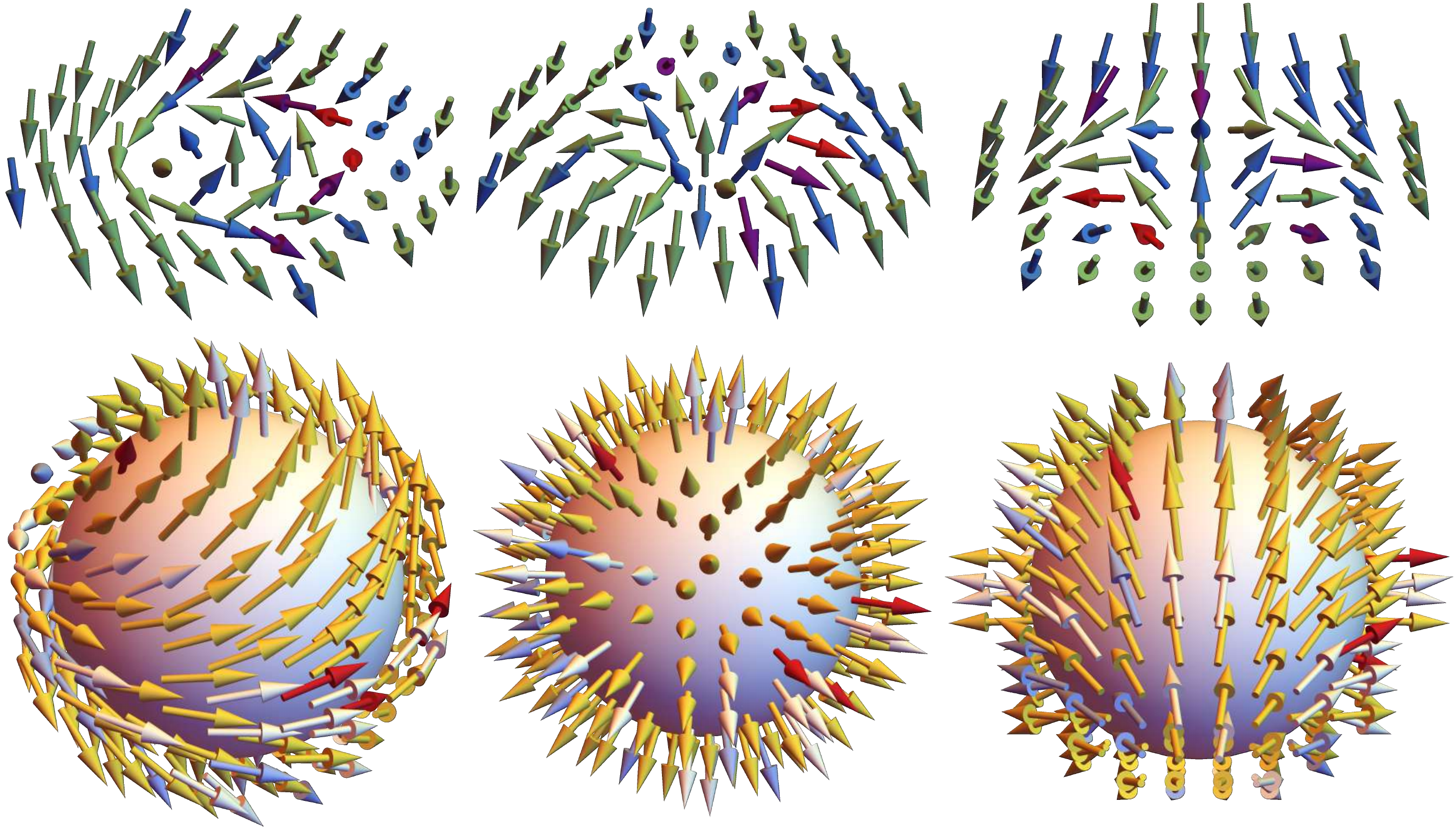}
\end{center}
\caption{\tiny Top: Bloch (left) and N\'eel (middle) skyrmions with topological charge $Q=1$ and polarity $p=1$. Antiskyrmion (right) with topological charge $Q=-1$ and $p=1$. Bottom: For N\'eel skyrmion (left), Bloch skyrmion (middle), and  antiskyrmion (right) the moments wrap around a unit sphere upon application of stereographic projection.}\label{fig1}
\end{figure}

A magnetic skyrmion is a non-collinear configuration of magnetic moments with a whirling magnetic structure (see Figure \ref{fig1}). Magnetic skyrmions represent a (2+1)-dimensional analog of the Skyrme model \citep{SkyrmePotRSoLSA1961}, which is a (3+1)-dimensional theory. Thus, magnetic skyrmions are often referred to as baby-skyrmions. From a topological point of view, a magnetic skyrmion is described by an integer invariant, referred to as topological charge, that arises in the map from the physical two-dimensional space to the target space $S_2$. The formula for topological charge is: 

\begin{align}
Q = \frac{1}{4 \pi}\int d^2 r (\partial_x \boldsymbol m \times \partial_y \boldsymbol m)\cdot \boldsymbol m ,
\end{align}
where $\boldsymbol m$ is a unit vector pointing in the direction of the magnetization. The topological charge describes how many times magnetic moments wrap around a unit sphere in the mapping (see Figure \ref{fig1}). Erasing a skyrmion requires globally modifying the system and, as a result, skyrmions possess topological protection. Initially, magnetic skyrmions of Bloch type (see Figure \ref{fig1}) \cite{Bogdanov.HubertJMMM1994,Roesler.Bogdanov.eaN2006,Muehlbauer.Binz.eaS2009,Yu.Onose.eaN2010} were discovered in chiral B20 compounds such as  MnSi, FeGe, and Fe$_{1-x}$Co$_x$Si in which spin-orbit interactions and the absence of the center of inversion lead to appearance of  the Dzyaloshinskii-Moriya interaction (DMI) \cite{DzyaloshinskyJPCS1958,MoriyaPR1960}. Sufficiently strong DMI can then lead to the formation of isolated skyrmions or even skyrmion lattices. 

Skyrmions stabilized by DMI are commonly referred to as Bloch- and N\'eel-type skyrmions (see Figure \ref{fig1}). Helicity and polarity are also used to describe skyrmions. Helicity can be defined as the angle of the global rotation around the z-axis that relates various skyrmions to the N\'eel skyrmion. For the N\'eel skyrmion, helicity is zero. Polarity describes whether the magnetization points in the positive ($p=1$) or negative ($p=-1$) z-direction at the center of the skyrmion \cite{Zheng.LiPRL2017}. For Bloch and N\'eel skyrmions, the topological charge and polarity are equal ($Q=p$). The difference in helicity distinguishes Bloch and N\'eel skyrmions from one another. On the other hand, magnetic textures stabilized by DMI can also have opposite topological charge and polarity ($Q=-p$). Such magnetic textures are referred to as \textit{antiskyrmions}. Antiskyrmions (see Figure \ref{fig1}) can be stabilized by bulk DMI with lower symmetry as was first predicted \cite{Bogdanov.YablonskiiJETP1989,Bogdanov.Roesler.eaPRB2002} and later realized experimentally at room temperatures in Heusler compounds with $D_{2d}$ symmetry \cite{Nayak.Kumar.eaN2017}.
It has also been predicted that interfacial DMI with $C_{2v}$ symmetry can lead to formation of antiskyrmions in ultrathin magnetic films \cite{Guengoerdue.Nepal.eaPRB2016,Hoffmann.Zimmermann.eaNC2017}. In general, various types of skyrmions and antiskyrmions, as shown in Figure \ref{figadd}, can be stabilized by changing the form of DMI tensor \cite{Guengoerdue.Nepal.eaPRB2016}. Anisotropic interfacial DMI with $C_{2v}$ symmetry has recently been realized in epitaxial Au/Co/W magnetic films \cite{Camosi.Rohart.eaPRB2017}.

\begin{figure}[h!]
\begin{center}
\includegraphics[width=0.3\columnwidth]{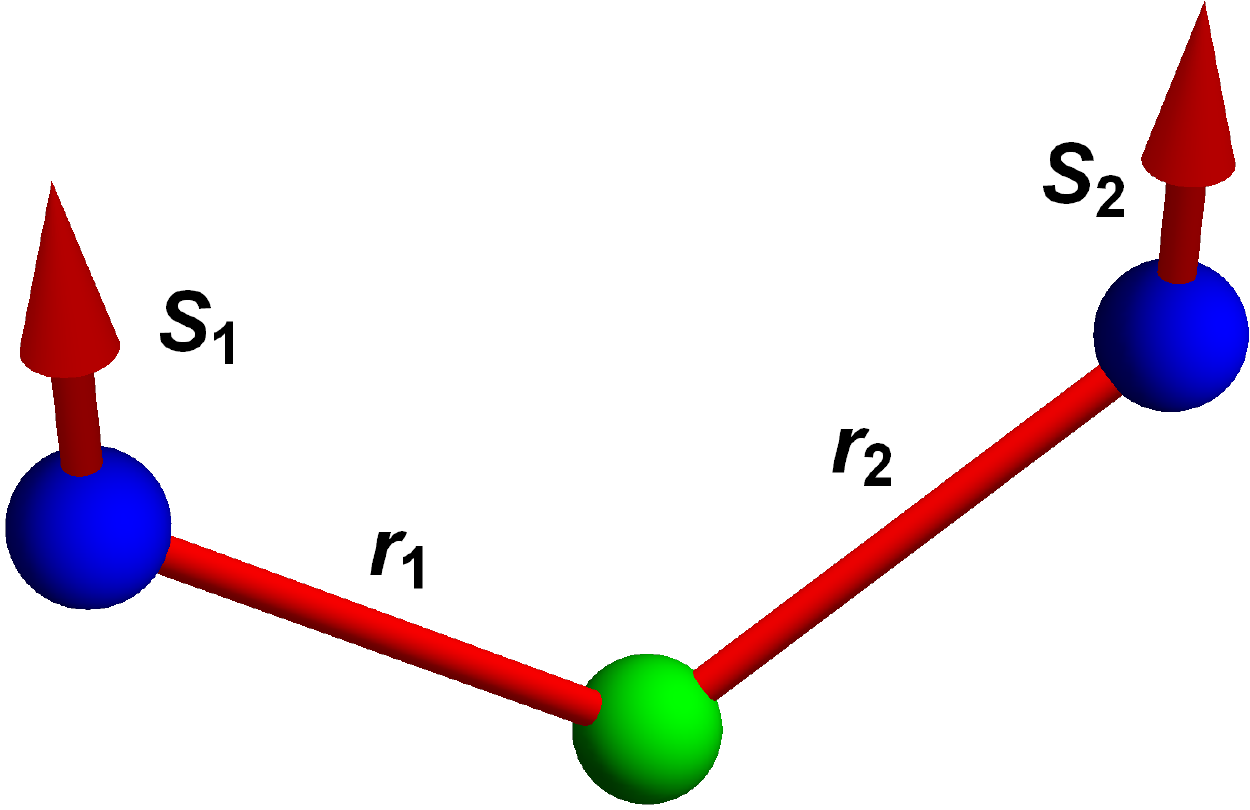}
\end{center}
\caption{\tiny Interaction between two neighbouring spins is mediated by the non-magnetic atom. The direction of DMI vector resulting from such interaction is defined by the bond vectors, i.e., $\boldsymbol {\mathcal D}_{12} \propto \boldsymbol r_1 \times \boldsymbol r_2$. }\label{fig2}
\end{figure}

\begin{figure}[h!]
\begin{center}
$Q=1$

\includegraphics[width=0.15\columnwidth]{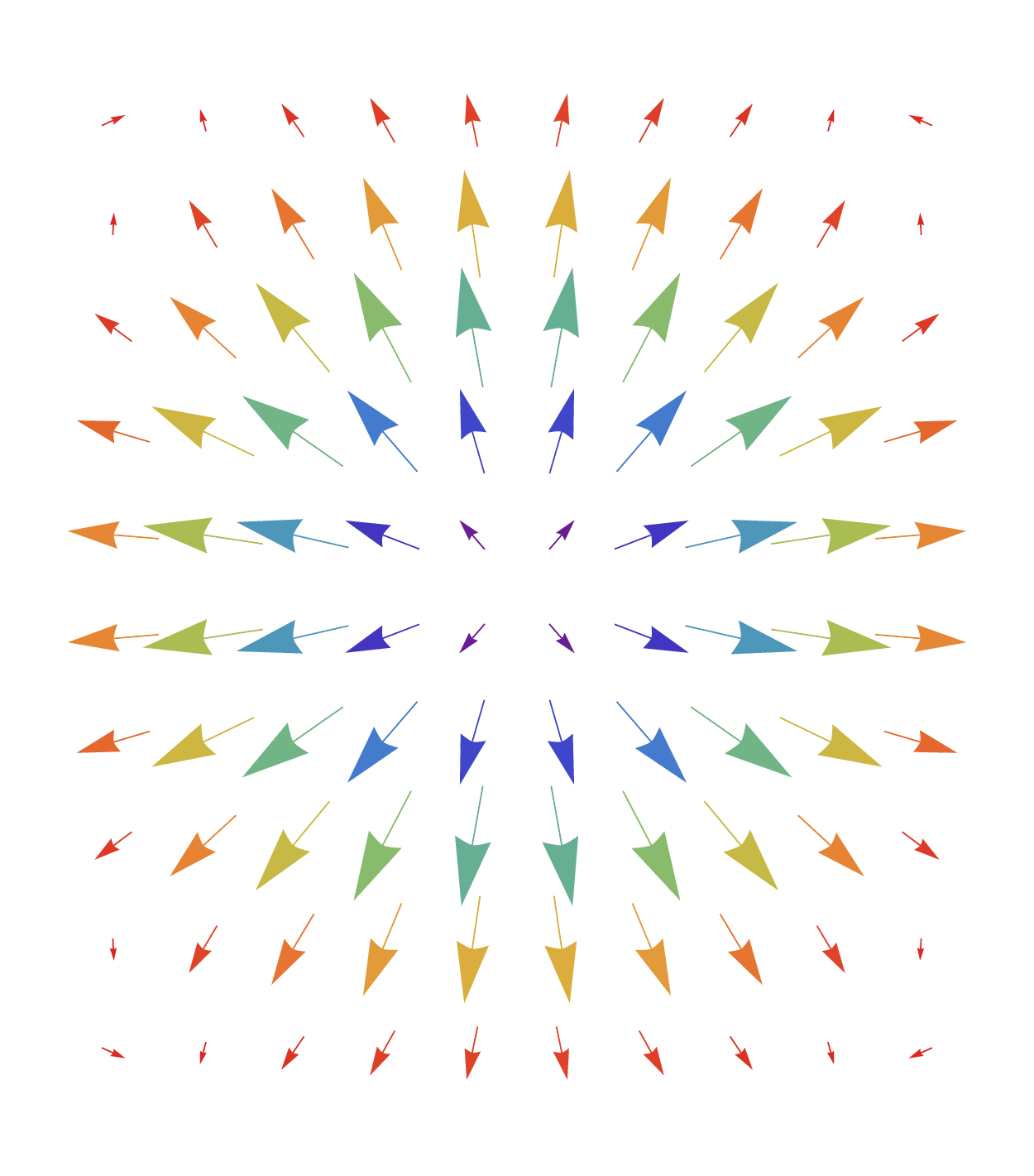}
\includegraphics[width=0.15\columnwidth]{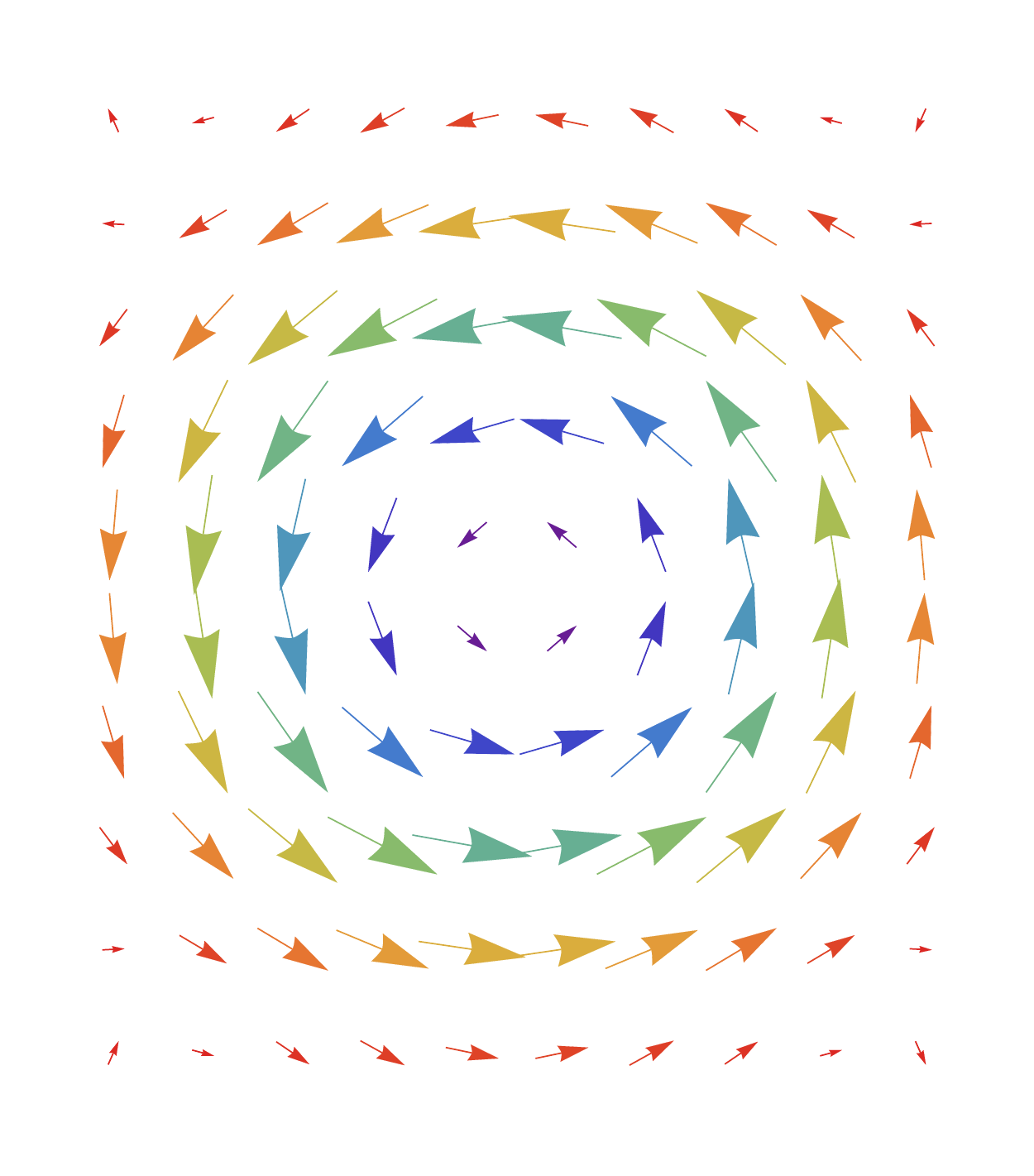}
\includegraphics[width=0.15\columnwidth]{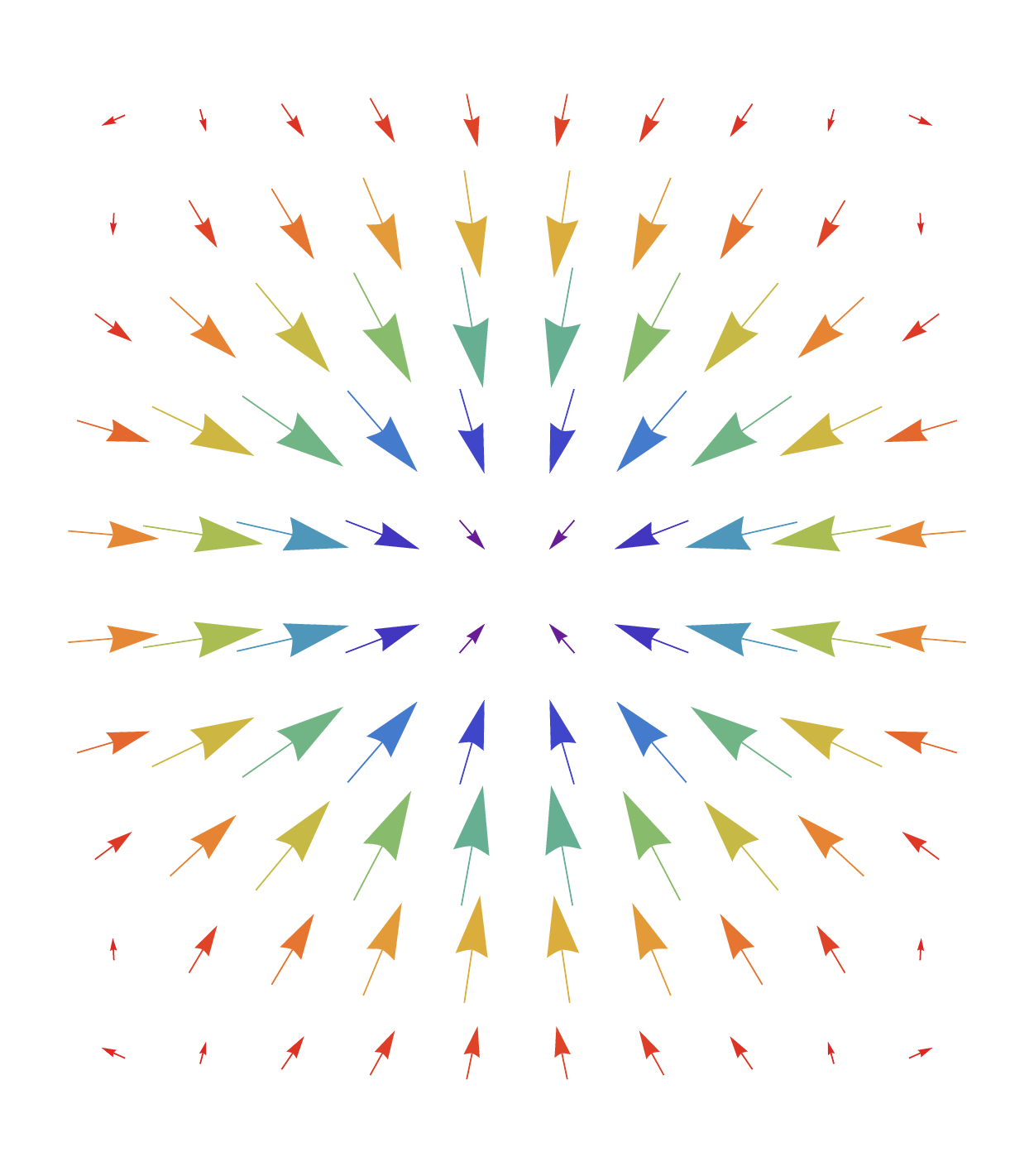}
\includegraphics[width=0.15\columnwidth]{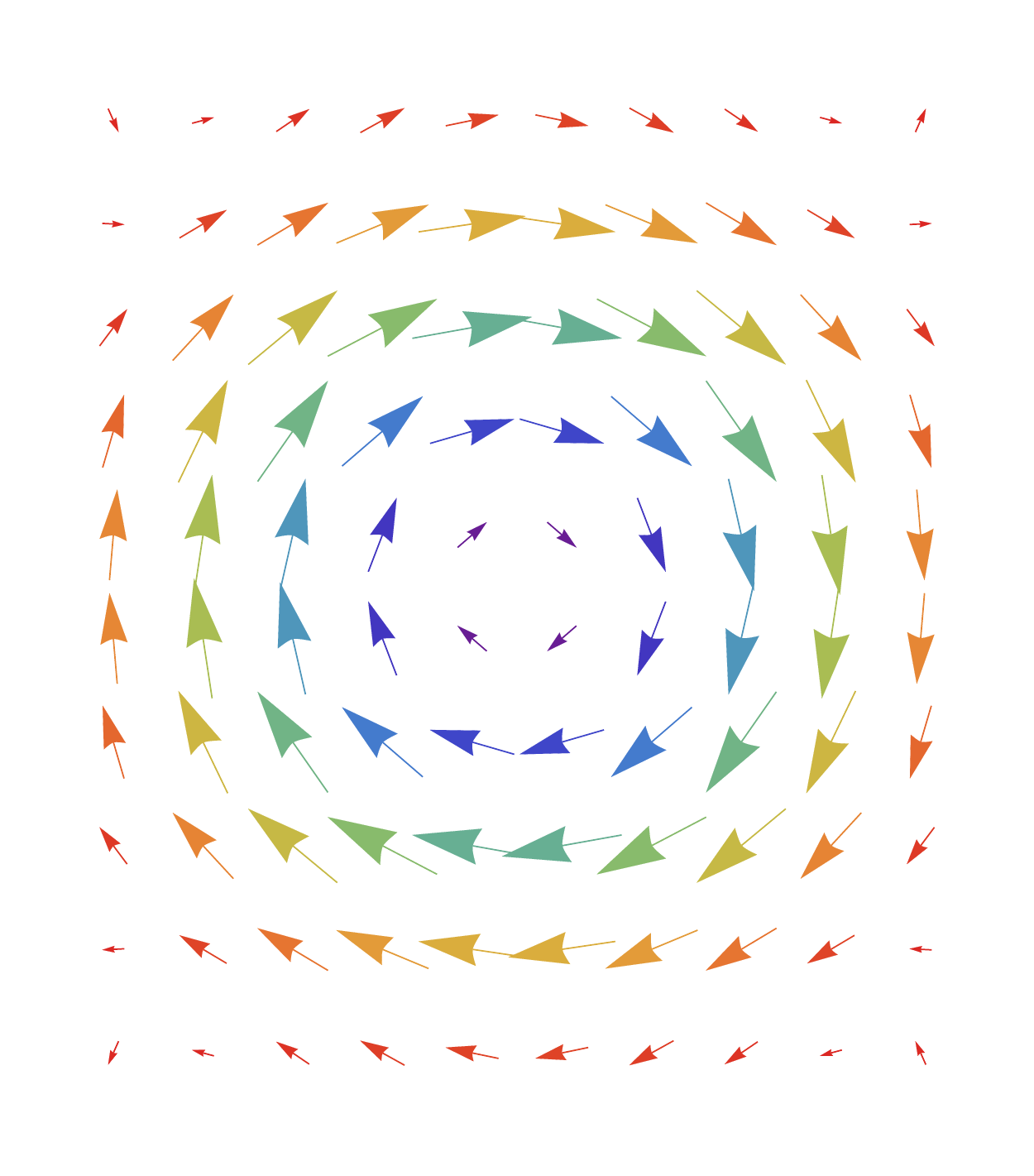}

$\quad\gamma=0\quad\quad\quad\quad \gamma=\frac{\pi}{2}\quad\quad\quad\quad \gamma=\pi\quad\quad\quad\quad \gamma=-\frac{\pi}{2}$
\end{center}
\vspace{10pt}
\begin{center}
$Q=-1$

\includegraphics[width=0.15\columnwidth]{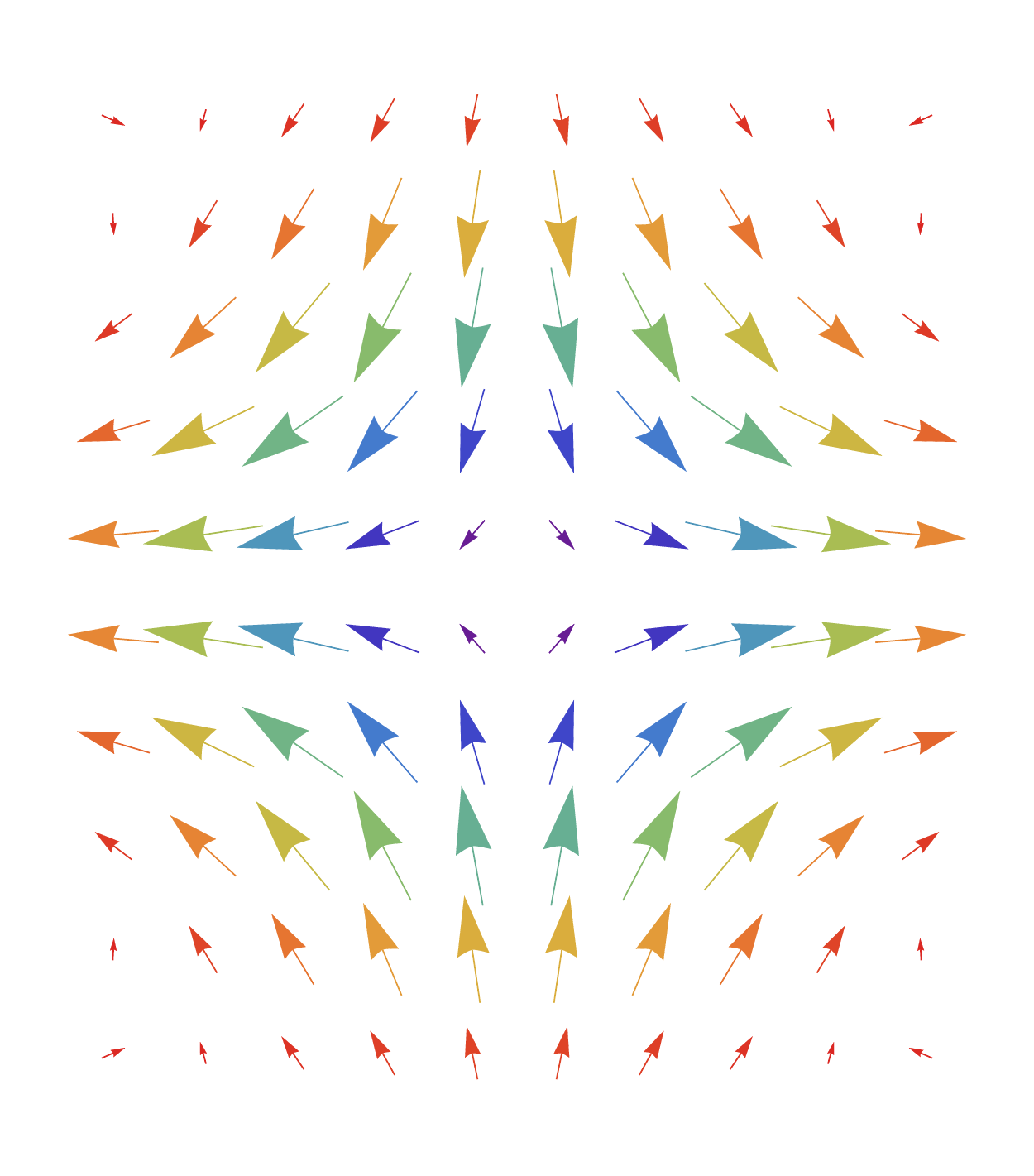}
\includegraphics[width=0.15\columnwidth]{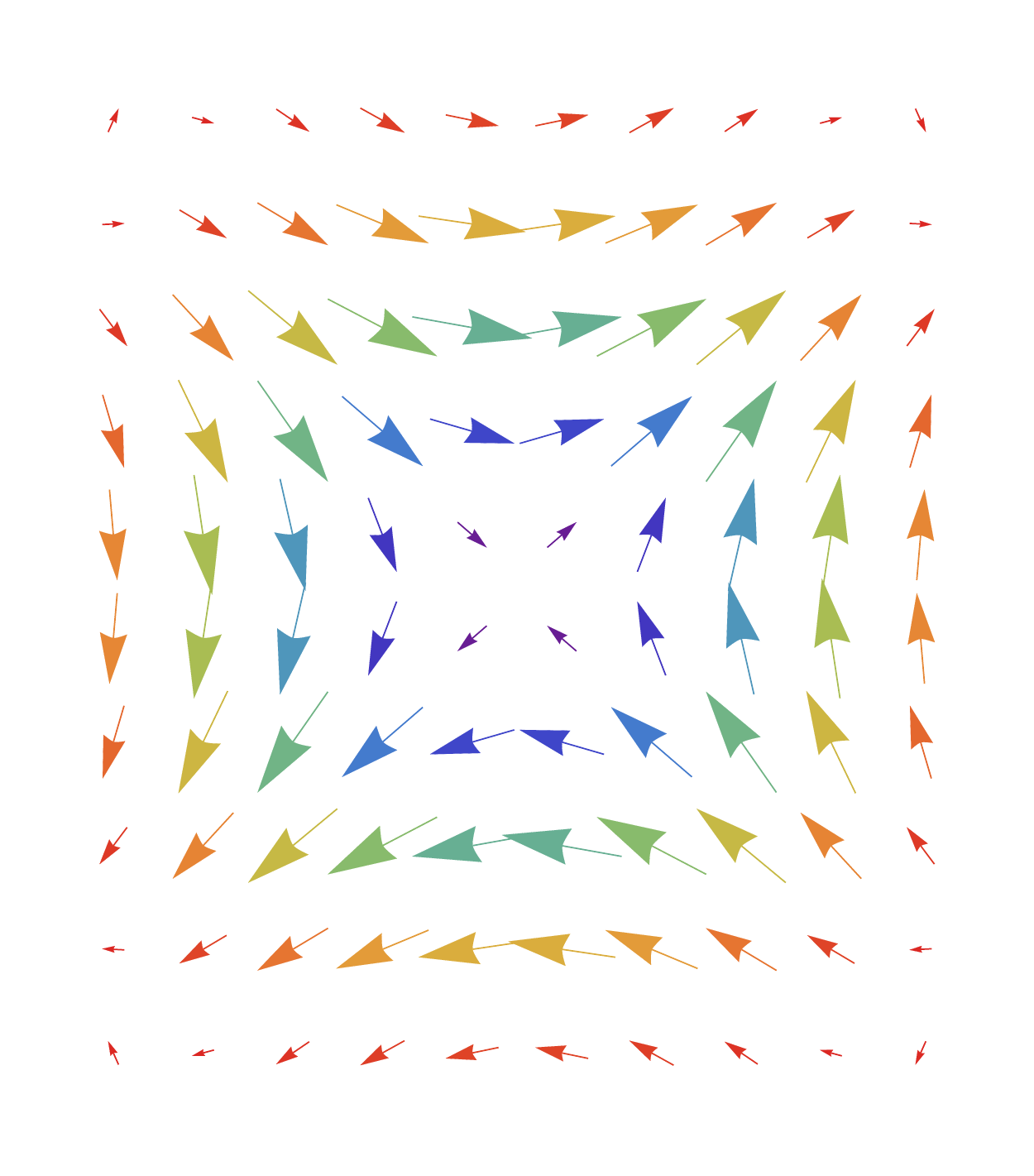}
\includegraphics[width=0.15\columnwidth]{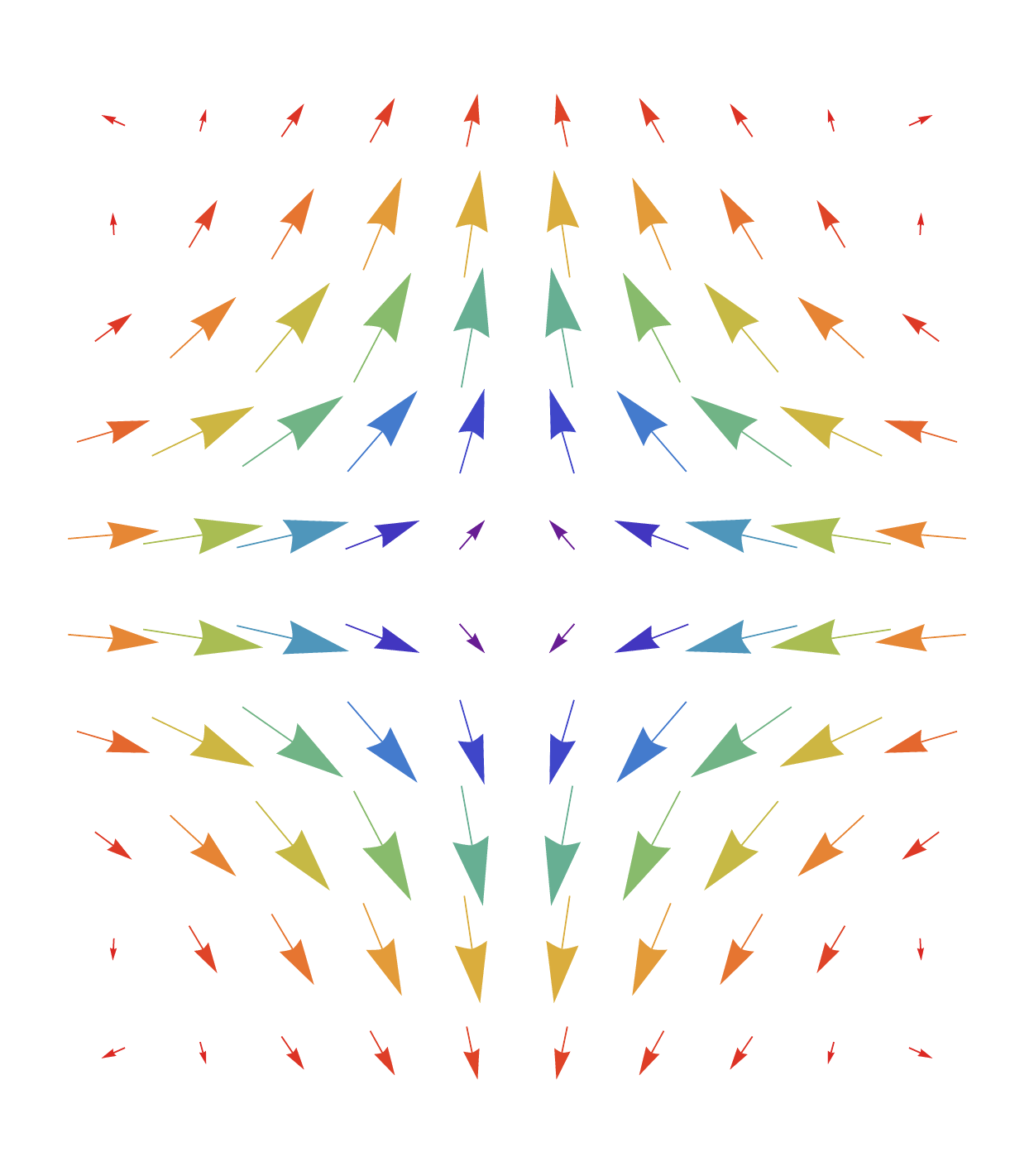}
\includegraphics[width=0.15\columnwidth]{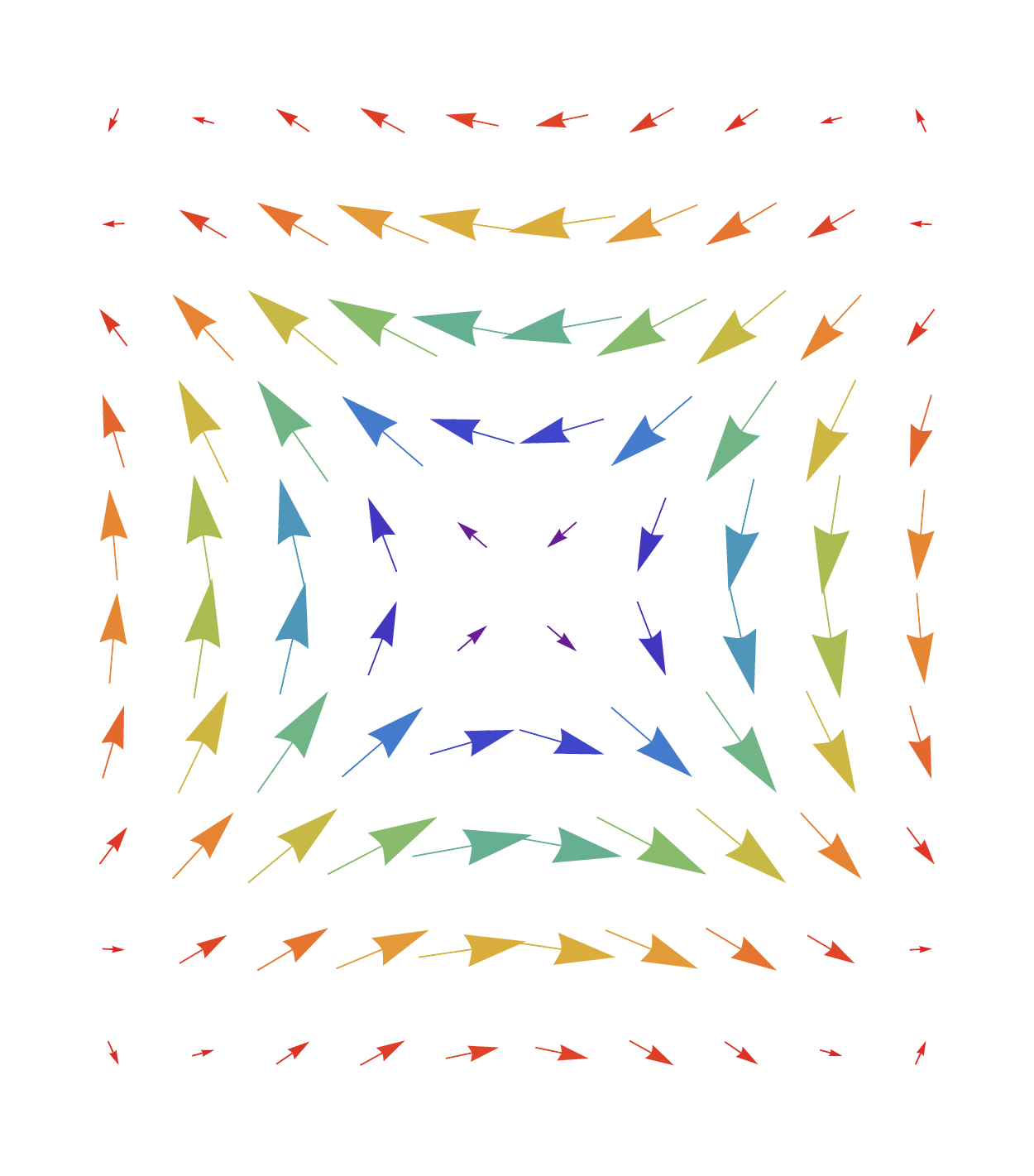}

$\quad\gamma=0\quad\quad\quad\quad \gamma=\frac{\pi}{2}\quad\quad\quad\quad \gamma=\pi\quad\quad\quad\quad  \gamma=-\frac{\pi}{2}$
\end{center}
\caption{\tiny Magnetization configuration of skyrmions ($Q=1$) and antiskyrmions ($Q=-1$) with fixed polarity ($p=1$) and varying helicity. The length and direction of the arrows represent the in-plane component of $\boldsymbol m$, and the color indicates $m_z$. A N\'eel skyrmion with $Q=1$ and helicity $\gamma=0$ corresponds to DMI with $C_{\infty v}$ symmetry. A Bloch skyrmion with $Q=1$ and helicity $\gamma=\pi/2$ corresponds to DMI with $SO(3)$ symmetry. Antiskyrmion with $Q=-1$ and helicity $\gamma=0$ corresponds to DMI with $D_{2d}$ symmetry. An arbitrary helicity $\gamma$ can be obtained via the global rotation arounds the $z$-axis by an angle $\gamma$. Antiskyrmions can be obtained from skyrmions by a reflection with respect to a plane perpendicular to the plane of the magnet. Figure is adapted with permission from Ref.~\cite{Guengoerdue.Nepal.eaPRB2016}.}\label{figadd}
\end{figure}

Microscopically, DMI arises when the interaction of two magnetic atoms is mediated by a non-magnetic atom via the superexchange or double-exchange mechanisms (see Figure \ref{fig2}). In the absence of an inversion center, a non-collinear configuration of magnetic moments is preferred by DMI energy:
\begin{equation}
H_{DMI}=\boldsymbol {\mathcal D}_{12}\cdot (\boldsymbol S_1 \times \boldsymbol S_2),
\end{equation}      
where $\boldsymbol S_1$ and $\boldsymbol S_2$ describe the spins and the DMI vector $\boldsymbol {\mathcal D}_{12} \propto \boldsymbol r_1 \times \boldsymbol r_2$ (see Figure \ref{fig2}). The strength of DMI is proportional to the strength of the spin-orbit interaction, which is expected to scale with the fourth power of the atomic number. However, in some cases the particular form of the band structure and/or effects related to charge transfer can influence the strength of the spin-orbit interaction. Particularly strong spin-orbit interaction and DMI can arise at interfaces between magnetic films and non-magnetic metals where the $3d$ orbitals of magnetic atoms interact with $5d$ orbitals of the heavy metal \cite{Fert.Cros.eaNN2013}. N\'eel type-skyrmions stabilized by interfacial DMI have been obtained at low temperatures in epitaxially grown Fe and PdFe magnetic layers on Ir \cite{Heinze.Bergmann.eaNP2011,Romming.Hanneken.eaS2013}. Another approach is to stack magnetic and non-magnetic layers in such a way that additive interfacial DMI leads to the formation of N\'eel skyrmions. Such an approach leads to the formation of room temperature skyrmions in magnetic layers (e.g., Co) sandwiched between two different non-magnetic layers (e.g., Ir and Pt) \cite{Moreau-Luchaire.Moutafis.eaNN2016,Woo.Litzius.eaNM2016,Boulle.Vogel.eaNN2016,Soumyanarayanan.Raju.eaNM2017}. 

Non-collinear magnetic textures can also arise due to dipolar interactions in the form of magnetic bubbles. Compared to skyrmions, magnetic bubbles have larger size and no definite chirality. Thus antiskyrmions can be realized in systems with dipolar interactions \cite{Koshibae.NagaosaNC2016}. Magnetic skyrmion bubbles are similar to magnetic bubbles but have definite chirality induced by DMI \cite{Jiang.Upadhyaya.eaS2015,Jiang.Zhang.eaNP2017,Yu.Upadhyaya.eaNL2017}. Skyrmion-like structures can be also realized in systems without DMI \cite{Polyakov:1975,volovik1977,dzyloshinskii1979}, e.g., it has been predicted that skyrmions of both chiralities can be stabilized in lattices with frustrated exchange interactions \cite{Okubo.Chung.eaPRL2012,Leonov.MostovoyNC2015,Zhang.Xia.eaNC2017}.

\section{Description of skyrmions and antiskyrmions}

Magnetic skyrmions in thin magnetic films can be well understood by considering a continuous model with the free energy density written for a two-dimensional ferromagnet well below the Curie temperature:
\begin{equation}\label{FreeE}
{\mathcal F}=A\left(\partial_{i}\boldsymbol m\right)^{2}-K m_{z}^{2}-H m_{z}+\boldsymbol D_j \cdot(\partial_j \boldsymbol m \times \boldsymbol m),
\end{equation}
where the free energy is $F=\int d^2 r \mathcal F$, we assume summation over repeated index $i=x,y$, and $\boldsymbol m$ is a unit vector along the magnetization direction. The first term in Eq.~(\ref{FreeE}) describes isotropic exchange with exchange stiffness $A$, the second term describes uniaxial anisotropy with strength $K$, the third term describes the Zeeman energy due to the external magnetic field $H_e$, $H\equiv\mu_{0}H_e M$ where $M$ is the magnetization, and the last term corresponds to DMI described by a general tensor $D_{ij}=(\boldsymbol D_j)_i$ \cite{Guengoerdue.Nepal.eaPRB2016}, where for a two-dimensional magnet $j$ is limited to $x$ and $y$. DMI is the most important term in Eq.~(\ref{FreeE}) for the formation of magnetic skyrmions. Modest DMI strength favors isolated metastable skyrmions \cite{Leonov.Monchesky.eaNJoP2016}, while strong DMI leads to condensation into a skyrmion lattice \cite{Banerjee.Rowland.eaPRX2014} (see discussion around Eq.~(\ref{crit})). 

The form of the DMI tensor $D_{ij}$ is determined by the crystallographic symmetry of the system \cite{Guengoerdue.Nepal.eaPRB2016}. In particular, non-zero elements of the DMI tensor are determined by relations:
\begin{equation} \label{eq:cryst}
D_{ij}=(\det \boldsymbol R^{(\alpha)}) R^{(\alpha)}_{il} R^{(\alpha)}_{jm} D_{lm},
\end{equation}
where $\boldsymbol R^{(\alpha)}$ are generators of the point group corresponding to the crystallographic symmetry, $\alpha=1,2,\ldots$, and the summation over repeated indices $l$ and $m$ is assumed. Note that the constraints on the DMI tensor in Eq.~(\ref{eq:cryst}) can be equivalently expressed via Lifshitz invariants \cite{DzyaloshinskiiSPJ1964,Bogdanov.YablonskiiJETP1989,Bak.JensenJPCSSP1980}. 

A system invariant under $SO(3)$ rotations then results in $D_{ij} =D \delta_{ij}$, where $\delta_{ij}$ is the Kronecker delta. Such DMI stabilizes Bloch-type skyrmions (see Figure \ref{fig1}). A system with $C_{\infty v}$ symmetry is invariant under proper and improper rotations around the z axis and allows only two nonzero tensor coefficients $D_{12}=-D_{21}=D$. Such DMI stabilizes N\'eel-type skyrmions (see Figure \ref{fig1}). Another important example arises for a system invariant under $D_{2d}$ symmetry, for which again only two nonzero tensor coefficients are allowed, i.e., $D_{12}=D_{21}=D$. The latter case realizes a system with \textit{antiskyrmions} (see Figure \ref{fig1}). Within a simple model given by Eq.~(\ref{FreeE}), all three examples given above are mathematically equivalent as they can be mapped to each other by a global spin rotation/reflection accompanied by an appropriate transformation of the DMI tensor \cite{Guengoerdue.Nepal.eaPRB2016}. Since the free energy does not change in such a mapping, one can expect that the same stability diagram will describe the above skyrmions and antiskyrmions (other examples of equivalent skyrmions and antiskyrmions are shown in Figure \ref{figadd}) \cite{Guengoerdue.Nepal.eaPRB2016}. 
This equivalence is no longer valid in the presence of dipolar interactions \cite{Camosi.Rougemaille.eaPRB2018} or more complicated magnetocrystalline anisotropy. In fact, antiskyrmions offer increased stability due to the presence of dipolar interactions \cite{Camosi.Rougemaille.eaPRB2018}.

\begin{figure}[h!]
\begin{center}
\tiny
\begin{overpic}[width=0.46\columnwidth]{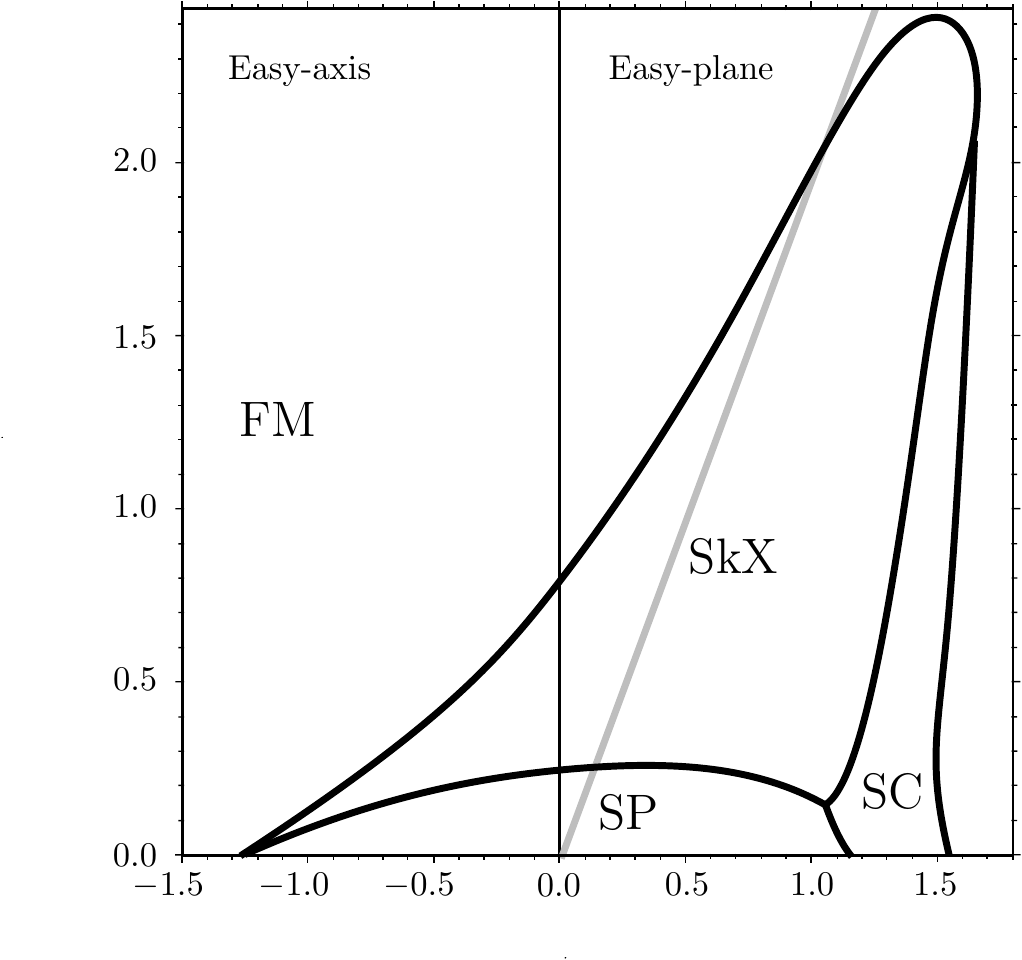}
\put(110,3){\colorbox{white}{$\quad\quad$}}
\put(115,3){$\frac{2A K}{D^2}$}
\put(0,119){\colorbox{white}{$\quad\quad$}}
\put(3,115){$\frac{2A H}{D^2}$}
\end{overpic}
\end{center}
\caption{\tiny  Zero temperature phase diagram for the three cases corresponding to $SO(3)$, $C_{\infty v}$, and $D_{2d}$ symmetries is obtained by numerically solving the LLG equation. The axes correspond to the dimensionless magnetic field and dimensionless uniaxial anisotropy parameter (see Eq.~(\ref{FreeE})). The gray line separates the aligned and the tilted regions of the FM phase. This phase is taken over by the hexagonal skyrmion lattcie (SkX), spiral (SP), and the square cell skyrmion lattice (SC) phases in the regions defined by the bold lines. Figure is adapted with permission from Ref.~\cite{Guengoerdue.Nepal.eaPRB2016}.}\label{fig3}
\end{figure}
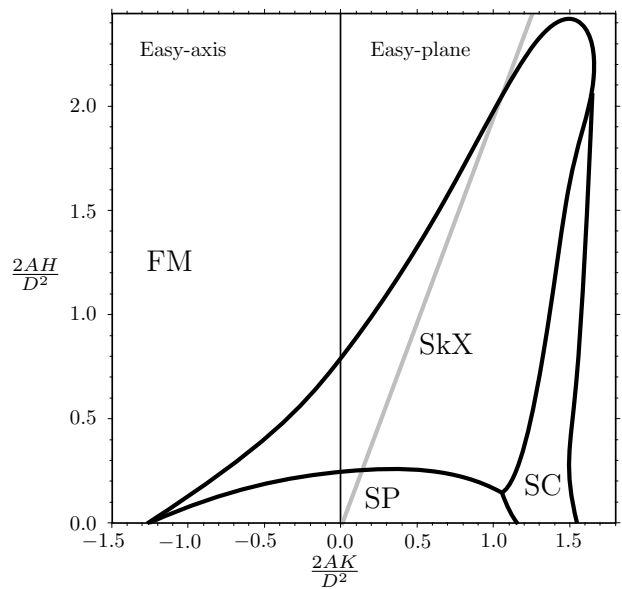

The parameters $A$, $K$, $H$, and $D$ in the above examples enter the free energy density (\ref{FreeE}), and they determine whether magnetic skyrmions or antiskyrmions can be present in a system. Minimization of the free energy corresponding to Eq.~(\ref{FreeE}) leads to the phase diagram shown in Figure \ref{fig3}, where the phase boundaries separate the cycloid or spiral phase (SP), the hexagonal skyrmion lattice (SkX), the square cell skyrmion lattice (SC), and the ferromagnetic phase (FM) \cite{Guengoerdue.Nepal.eaPRB2016}. It is convenient to introduce a critical DMI: 
\begin{equation} D_c=4(AK)^{1/2}/\pi,\label{crit}
\end{equation} 
corresponding to the strength of DMI at which the formation of Dzyaloshinskii domain walls becomes energetically favorable \cite{Rohart.ThiavillePRB2013}. In an infinite sample, the transition from isolated skyrmions to a skyrmion lattice or cycloid phase happens in the vicinity of this critical DMI strength \cite{Banerjee.Rowland.eaPRX2014,Siemens.Zhang.eaNJP2016,Guengoerdue.Nepal.eaPRB2016} (see Figure \ref{fig3}). 
The magnetic skyrmion size changes substantially as one varies the strength of DMI. At $D<D_c$, the skyrmion size has been calculated analytically, $R_s\approx\Delta/\sqrt{2(1-D/D_c)}$, where $\Delta=\sqrt{A/K}$ \cite{Rohart.ThiavillePRB2013}. The effects related to finite temperature and dipolar interactions modify this behavior, especially close to the divergence when $R_s \rightarrow \infty$ at $D=D_c$ \cite{Boulle.Vogel.eaNN2016}. At $D>D_c$, inside the skyrmion lattice phase the skyrmion lattice period can be estimated by the period of the equilibrium helix, $L_D= 4\pi A/D$ \cite{Banerjee.Rowland.eaPRX2014,McGrouther.Lamb.eaNJP2016,Guengoerdue.Nepal.eaPRB2016}. 
Note that the ballpark value for the critical DMI is $D_c\sim 4$ mJ/m$^2$ (we use parameters for Co/Pt multilayer \cite{Miron.Moore.eaNM2011}). Comparable DMI can be realized in magnetic layers sandwiched between non-magnetic layers. The skyrmion size can range from 8 nm at low temperatures ($<$30K) \cite{Bode.Heide.eaN2007} to 50 nm at room temperature \cite{Moreau-Luchaire.Moutafis.eaNN2016,Woo.Litzius.eaNM2016,Boulle.Vogel.eaNN2016}.

In principle, the stability of a skyrmion can be hindered by finite temperature. In studies of magnetic memories, this question is usually answered by studying the Arrhenius law of escape from a particular state \cite{LangerAP1969}. Recent studies based on harmonic transition state theory confirm that the life-time of isolated skyrmions is sufficient for realizations of magnetic memories \cite{Bessarab.Mueller.eaSR2018}.

\section{Dynamics of skyrmions and antiskyrmions}

A design of skyrmion-based memory device \cite{Fert.Cros.eaNN2013} closely resembles the race-track memory \cite{Hayashi.Thomas.eaS2008}. One of the challenges is to achieve efficient control of the skyrmion dynamics. Skyrmion dynamics can be induced by spin transfer torques (STT) \cite{Jonietz.Muehlbauer.eaS2010,Yu.Kanazawa.eaNC2012,Sampaio.Cros.eaNN2013,Iwasaki.Mochizuki.eaNN2013}, spin-orbit torques (SOT) (often refered to as spin Hall effect torques) \cite{Sampaio.Cros.eaNN2013,Woo.Song.eaNC2017}, voltage-controlled magnetic anisotropy \cite{Kang.Huang.eaSR2016}, surface acoustic waves \cite{Nepal.Guengoerdue.eaAPL2018}, temperature gradients \cite{Kong.ZangPRL2013,Lin.Batista.eaPRL2014,KovalevPRB2014,Mochizuki.Yu.eaNM2014}, and other mechanisms. 

Here, we discuss (anti)skyrmion dynamics due to STT and SOT induced by in-plane charge currents \cite{Huang.Zhou.eaPRB2017,Zhang.Xia.eaNC2017}. These two mechanisms differ as the former vanishes in the absence of magnetic textures while the latter does not. 
Both mechanisms can be included in the Landau-Lifshitz-Gilbert equation:

\begin{align}
s (1- \alpha \boldsymbol m \times) \dot {\boldsymbol m} =  \boldsymbol H_\text{eff}\times \boldsymbol m - (1 -\beta \boldsymbol m \times) (\boldsymbol j_s \cdot \boldsymbol  \nabla ) \boldsymbol m +\boldsymbol \tau_{so} ,
\label{eq:LLG}
\end{align}
where $s=M_s/\gamma$ is the spin density, $M_s$ is the saturation magnetization, $\gamma$ is (minus) the gyromagnetic ratio ($\gamma>0$ for electrons), $\alpha$ is the Gilbert damping, $H_\text{eff}=-\delta F/\delta \boldsymbol m$ is the effective field, $\beta$ is the factor describing non-adiabaticity, $\boldsymbol j_s$ is the in-plane spin current proportional to the charge current $\boldsymbol j$, and $\boldsymbol \tau_{so}$ is the spin-orbit torque \cite{Brataas.Kent.eaNM2012}. For the spin Hall contribution, $\boldsymbol \tau_{so}=(\hbar\theta_{SH}/2e t_f)\boldsymbol m\times[\boldsymbol m\times(\hat{z}\times\boldsymbol j)]$, where $\theta_{SH}$ is the spin Hall angle, $e$ is the electron charge, and $t_f$ is the thickness of the ferromagnetic layer. 
The Thiele approach \cite{ThielePRL1973} applied to Eq.~(\ref{eq:LLG}) leads to equations of motion describing (anti)skyrmion dynamics:

\begin{align}
  ( Q \hat z\times + \alpha \hat\eta) \boldsymbol v  =\boldsymbol F ,\label{force}
\end{align}
where $\hat \eta$ is the damping dyadic tensor and $\boldsymbol F=\boldsymbol F_{so}+\boldsymbol F_{st}$ is the total force acting on the (anti)skyrmion due to SOT and STT. The SOT contribution is $\boldsymbol F_{so}=\hat{\mathcal{B}}\cdot \boldsymbol j$ where the tensor $\hat{\mathcal{B}}$ is proportional to the spin Hall angle and is determined by the configuration of the (anti)skyrmion \cite{Tomasello.Martinez.eaSR2014,Huang.Zhou.eaPRB2017}. The STT contribution is $\boldsymbol F_{st}=( Q \hat z\times + \beta \hat\eta)\boldsymbol j_s$ \cite{Iwasaki.Mochizuki.eaNN2013}.  We obtain the (anti)skyrmion velocity in an infinite sample:

\begin{align}
v_x = \frac{ F_y Q + F_x \alpha \eta  }{ Q^2 + \alpha^2 \eta^2}, \quad v_y =   \frac{-F_x Q + F_y \alpha \eta  }{ Q^2 + \alpha^2 \eta^2},\label{eq:vel}
\end{align}
where for typical (anti)skyrmions $\eta$ is of the order of $1$. The velocity in Eq.~(\ref{eq:vel}) scales as $1/Q$. An interesting situation happens in a nanotrack geometry close to the edge. Due to the edge repulsion, an additional force term appears in Eq.~(\ref{eq:vel}), $v_y=0$, and the (anti)skyrmion velocity becomes $v_x = F_x/(\alpha \eta)$. As a result, a large (anti)skyrmion velocity is possible close to the edge due to SOT in systems with low Gilbert damping \cite{Sampaio.Cros.eaNN2013,Hrabec.Sampaio.eaNC2017}. 

From Eq.~(\ref{eq:vel}), it is clear that (anti)skyrmions will move along a current with an additional side motion, resulting in the (anti)skyrmion Hall effect with the Hall angle $\theta_H = \tan^{-1}(v_y / v_x)$. Antiskyrmions exhibit an anisotropic Hall angle dependent on the current direction due to SOT, as the tensor $\hat{\mathcal{B}}$ becomes anisotropic for an antiskyrmion profile. This is in contrast to the isotropic behavior of skyrmions (the principal values are given by $\mathcal{B}_{xx}=-\mathcal{B}_{yy}$ for antiskyrmions with $\gamma=0$ in Figure \ref{figadd}, and $\mathcal{B}_{xx}=\mathcal{B}_{yy}$ for N\'eel skyrmions) \cite{Huang.Zhou.eaPRB2017}. It is also possible to completely suppress the antiskyrmion Hall effect by properly choosing the direction of the charge current or by coupling skyrmions and antiskyrmions in a multilayer stack \cite{Huang.Zhou.eaPRB2017}. Note that antiferromagnetic skyrmions \cite{Barker.TretiakovPRL2016} or synthetic antiferromagnetic skyrmions in antiferromagnetically coupled layers \cite{Zhang.Ezawa.eaPRB2016,Woo.Song.eaNC2018} also exhibit a vanishing Hall angle. This behavior can be useful for realizations of magnetic memories.

Skyrmion dynamics consistent with Eq.~(\ref{eq:vel}) have been observed in magnetic multilayers, e.g.,
in Pt/Co/Ta \cite{Woo.Litzius.eaNM2016}, Ta/CoFeB/TaO$_x$ \cite{Jiang.Upadhyaya.eaS2015,Jiang.Zhang.eaNP2017}, (Pt/CoFeB/MgO)$_{15}$ \cite{Litzius.Lemesh.eaNP2017}, and  (Pt/Co/Ir)$_{10}$ \cite{Legrand.Maccariello.eaNL2017}. SOT and STT contribute to Eq.~(\ref{force}) differently. The major component of SOT force pushes skyrmion in the direction of the current flow while the major component of STT pushes skyrmion in the transverse direction. It follows from Eq.~(\ref{eq:vel}) that in the presence of the edge repulsion the velocity due to SOT, $v_x \propto F_{SOT}/\alpha$, is expected to be larger than the velocity due to STT, $v_x \propto F_{STT}$. Theoretical modeling \cite{Sampaio.Cros.eaNN2013} predicts that the velocity due to SOT is 60 m/s for the current $2\times 10^{11}$Am$^{-2}$. Similar modelling of the motion induced by STT results in much smaller velocity of 8 m/s for the same current. It has also been confirmed experimentally that in a sufficiently large sample the sign of the transverse response is determined by the sign of the spin Hall angle and the topological charge. Some aspects of skyrmion dynamics are still puzzling, i.e., the dependence of the skyrmion Hall angle on the velocity and the dependence of the velocity on the skyrmion size. Variations in the skyrmion Hall angle have been attributed to the presence of pinning sites \cite{Reichhardt.Ray.eaPRL2015}. In principle, the presence of pinning sites could be problematic for realizations of magnetic memories \cite{Gross.Akhtar.eaPRM2018}. Nevertheless, the presence of the skyrmion Magnus force induced by pinning sites results in significant reduction of the depinning threshold current density \cite{Lin.Reichhardt.eaPRB2013,Reichhardt.Ray.eaPRL2015}, e.g., by several orders of magnitude if compared to domain wall dynamics. Experimentally, the depinning threshold current density can be as low as $10^{6}$Am$^{-2}$ \cite{Yu.Kanazawa.eaNC2012}. 

\section{Writing, erasing, and detecting}

Interfacial skyrmions in Fe on Ir have been controllably written and erased by STT induced by an STM tip at $4.2$K in the presence of high magnetic fields \cite{Romming.Hanneken.eaS2013}. For practical applications it is necessary to write and erase skyrmions at room temperature in the presence of weak or, preferably, no magnetic fields.  STT has been demonstrated to generate skyrmions in a confined geometry of a Pt/Co nanodot \cite{Sampaio.Cros.eaNN2013}. A SOT mechanism in which a pair of domain walls is converted into a skyrmion has been demonstrated numerically to require relatively large current densities \cite{Zhou.EzawaNC2014}. At room temperature, SOT has been used to create elongated chiral domains that, under application of inhomogeneous current, break into skyrmion bubbles \cite{Jiang.Upadhyaya.eaS2015}. In this experiment, the inhomogeneous current of relatively low magnitude, $5\times 10^{8}$~Am$^{-2}$, blows chiral domains into bubbles in analogy with how the Rayleigh-Plateau instabilities lead to formation of bubbles in fluid flows.  This process has been reproduced by micromagnetic simulations \cite{LinPRB2016}. Detailed analysis of topological charge density revealed that in the process of creating a skyrmion by SOT an unstable antiskyrmion can also be created for a period of time comparable to 1 ns \cite{Liu.Yan.eaAPL2016}. As an antiskyrmion is not stable in the presence of interfacial DMI ($C_{\infty v}$ case), it eventually annihilates due to the presence of the Gilbert damping \cite{Stier.Haeusler.eaPRL2017}. Thus, detailed studies of antiskyrmions can help in realizing new ways of writing and erasing skyrmions. Generation of skyrmions by charge currents has also been demonstrated for magnetic multilayer stacks \cite{Legrand.Maccariello.eaNL2017}, as well as for symmetric bilayers hosting pairs of skyrmions with opposite chirality \cite{Hrabec.Sampaio.eaNC2017}. Other approaches include laser-induced generation of topological defects \cite{Finazzi.Savoini.eaPRL2013,Berruto.Madan.eaPRL2018}. The feasibility of such an approach has been confimed theoretically \cite{Koshibae.NagaosaNC2014,Fujita.SatoPRB2017,Fujita.SatoPRB2017a,Yudin.Gulevich.eaPRL2017}.

For device application, it is preferable to detect the presence of a skyrmion electrically. Several transport techniques have been suggested. Non-collinear magnetoresistance can be used to detect a skyrmion where changes in the band structure induced by non-collinear moments are detected by STM \cite{Hanneken.Otte.eaNN2015}. However, only non-collinearity is being detected and no information about the topological structure of a skyrmion is obtained. Measuring the z-component of magnetization is possible by employing the anomalous Hall effect \cite{Nagaosa.Sinova.eaRMP2010}. This technique has been used to detect a single skyrmion \cite{Maccariello.Legrand.eaNN2018}. Measurement of the topological Hall effect can directly reveal the topological nature of a skyrmion as the effect originates in the fictitious magnetic field proportional to the topological charge.  The presence of skyrmions has been detected by measurements of the topological Hall effect in non-cetrosymmetric bulk materials in the B20 group \cite{Lee.Kang.eaPRL2009,Neubauer.Pfleiderer.eaPRL2009}. In principle, antiskyrmions \cite{Nayak.Kumar.eaN2017} should also exhibit the topological Hall effect but with the reversed sign due to the opposite topological charge. On the other hand, in magnetic mulitlayers the topological Hall effect is expected to be much smaller than the anomalous Hall contribution \cite{Maccariello.Legrand.eaNN2018}.

\section{Conclusions and Outlook}

In this article, we review recent progress in the field of skyrmionics -- a field concerned with studies of tiny whirls of magnetic configurations for novel memory and logic applications. A particular emphasis has been given to antiskyrmions. These, similar to skyrmions, are particle-like structures with topological protection. Compared to skyrmions, antiskyrmions have opposite topological charge and are anisotropic. Recent experimental observation of antiskyrmions at room temperature \cite{Nayak.Kumar.eaN2017} encourages further studies of transport and dynamical properties such as the topological Hall effect, antiskyrmion Hall effect, antiskyrmion nucleation and stability, and others. On the other hand, realization of antiskyrmions requires careful material engineering which should also be addressed in future studies. All in all, antiskyrmions offer some advantages over skyrmions as they can be driven without the Hall-like motion, offer increased stability due to dipolar interactions, and can be realized above room temperature. A possibility to annihilate a skyrmion-antiskyrmion pair can lead to new concepts of logic devices \cite{Zhang.Ezawa.eaSR2015}. 
Reservoir computing is another application in which skyrmions could prove highly useful. The anisotropic magnetoresistance (AMR) and the motion and deformation of magnetic texture caused by the interaction of a voltage-induced electric current and a single skyrmion can provide a nonlinear relationship between voltage and current, which is a key ingredient for reservoir computing \cite{Prychynenko.Sitte.eaPRA2018}. Skyrmion fabrics, which may include skyrmions, antiskyrmions, skyrmion crystal structure, and domain walls, represent a potential way to implement skyrmion-based reservoir computing since they fulfill the requirements of an Echo State Network (ESN)   \cite{Bourianoff.PinnaAIP2018}.
It is expected that, similar to skyrmions, antiskyrmions can potentially result in new device concepts for memory and neuromorphic computing applications \cite{Prychynenko.Sitte.eaPRA2018}. 

\section*{Conflict of Interest Statement}

The authors declare that the research was conducted in the absence of any commercial or financial relationships that could be construed as a potential conflict of interest.

\section*{Author Contributions}

AAK conceived the project presented in this mini review article. AAK and SS prepared the manuscript.

\section*{Funding}
This work was supported by the DOE Early Career Award DE-SC0014189.

\section*{Acknowledgments}
The topic editors are acknowledged for supporting this open-access publication.

\bibliographystyle{frontiersinHLTH&FPHY} 
\bibliography{test}

\newpage
\section*{Figure captions}


\end{document}